# Recent Advances in Design of Electrocatalysts for High-Current-Density Water Splitting

Yuting Luo[1], Zhiyuan Zhang[1], Manish Chhowalla[2,*], Bilu Liu[1,*]

1. Shenzhen Geim Graphene Center, Tsinghua-Berkeley Shenzhen Institute & Institute of Materials Research, Tsinghua Shenzhen International Graduate School, Tsinghua University, Shenzhen 518055, P. R. China.

2. Materials Science and Metallurgy, University of Cambridge, Cambridge CB3 0FS, UK

## Abstract

Electrochemical water splitting technology for producing "green hydrogen" is important for the global mission of carbon neutrality. Electrocatalysts with decent performance at high current densities play a central role in the industrial implementation of this technology. The field has advanced immensely in recent years, as witnessed by many types of catalysts have been designed and synthesized which work at industrially-relevant current densities (> 200 mA cm$^{-2}$). Note that the activity and stability of catalysts can be influenced by their local reaction environment, which are closely related to the current density. By discussing recent advances in this field, we summarize several key aspects that affect the catalytic performance for high-current-density electrocatalysis, including dimensionality of catalysts, surface chemistry, electron transport path, morphology, and catalyst-electrolyte interplay. We highlight the multiscale design strategy that considers these aspects comprehensively for developing high-current-density catalysts. We also put forward out perspectives on the future directions in this emerging field.

## 1. Introduction

Energy, water, and the environment are three of the top ten challenges faced by human beings, both now and in the next tens of years, as proposed by the late Nobel laureate Richard E. Smalley.[1] According to the International Energy Agency, world energy consumption grew to 9,938 Mtoe (million tons of oil equivalent) in 2018, of which about 70% was from fossil fuels, resulting in a record high $CO_2$ emission of over 33 gigatons.[2] The need to address the problems of environment and climate changes is driving a dramatic global transformation of energy systems. The electricity accounts for nearly 20% of the world's total energy consumption nowadays,[2] which is expected to overtake oil and coal and reach >30% in 2040.[3] The growing demand for renewable energy is a main driving force behind the rise electricity use, in which two-thirds of it is expected to be generated from renewable resources in 2040,[3] with a projected electricity price of less than 10 US cents per MWh by 2050.[4]

The need for renewable energy integrated with electricity generating systems is becoming more urgent than ever,[5] which requires the development of advanced energy conversion and storage technologies.[6] A sustainable way is to produce "green hydrogen" by electrochemical water splitting,[7, 8] coupled with electricity produced by renewable resources, as shown in Figure 1. Hydrogen is not only a promising alternative energy carrier to fossil fuels,[5] but also a crucial feedstock in industry for fertilizer production, petroleum refining, and hydrogenation. The main reactions involved in electrochemical water splitting include the hydrogen evolution reaction (HER) and oxygen evolution reaction (OER). In the context of global carbon neutrality, the importance of "green hydrogen" by electrochemical water splitting technology has arose massive attention not only by the scientific community but also by governments and industries around the world.

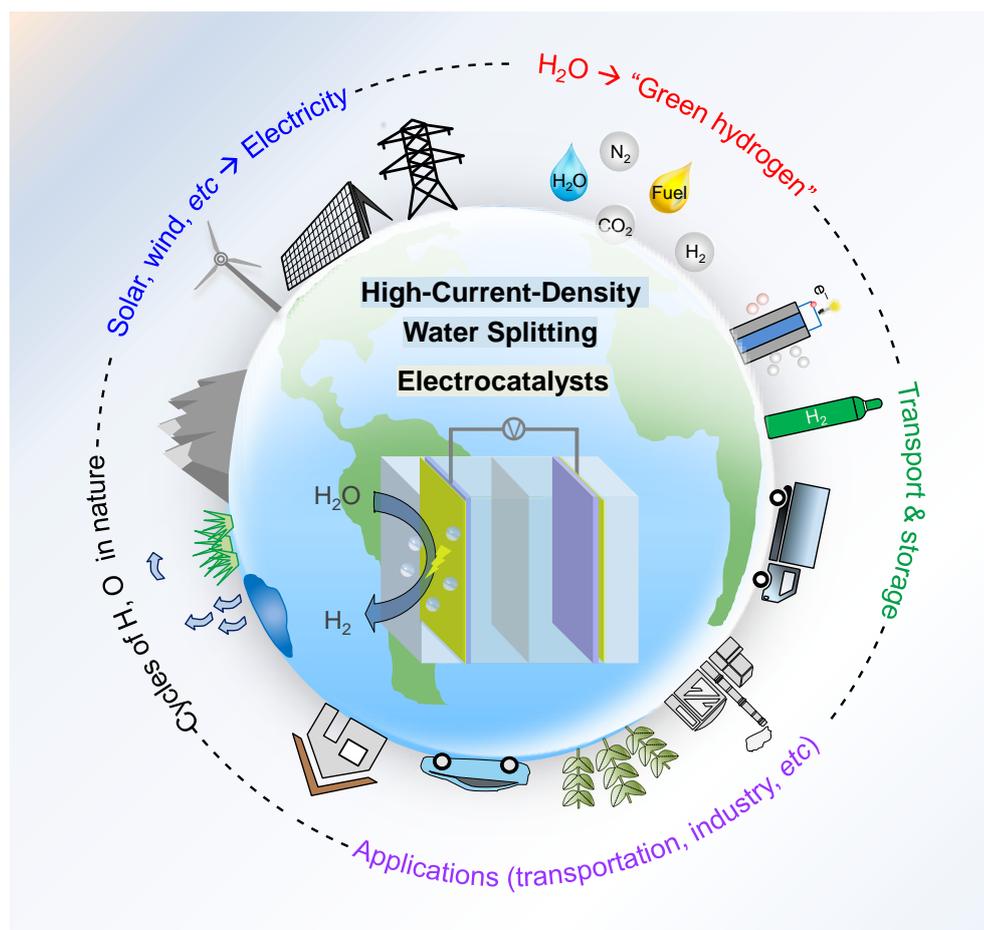

**Figure 1.** A schematic showing the crucial role of high-current-density (HCD) electrocatalysts for the production of "green hydrogen" by electrochemical water splitting technology coupled with renewable electricity.

For industrial use, developing electrocatalysts with a good performance under the industrially-relevant conditions including high current density (HCD), long working time, and demanded pressure and temperature, is crucial. The industrially-relevant current density is necessary because HCD means a high rate of hydrogen production, which can reduce capital expenditures and lead to a profitable hydrogen production. In this regard, many governments and organizations have proposed different technical targets at HCDs which are needed for different applications. For example, the current density requirement for the central proton exchange membrane (PEM) water electrolysis is 1500 mA cm$^{-2}$ at cell voltage of 1.75 V in 2015 and will reach 1600 mA cm$^{-2}$ at 1.66 V in 2040, according to technical targets from the U.S. Department of Energy (DOE).[9] The Fuel Cells and Hydrogen Joint Undertaking in Europe (FC HJU) proposes more ambiguous goals showing that a current density of 800 mA cm$^{-2}$ for alkaline water electrolysis and 2500 mA cm$^{-2}$ for PEM water electrolysis should be achieved in 2030.[10] Table 1 summarizes some future targets of performance metrics of water splitting technologies. These targets motivate studies of electrochemical water splitting under HCD conditions.

**Table 1.** Summary of the future performance targets for the high-current-density water electrolysis technologies.

| Source | Technology | Current density (mA cm$^{-2}$) | Voltage (V) | Energy efficiency (%) | Stability | Temperature (°C) | Pressure (atm) |
|---|---|---|---|---|---|---|---|
| U.S. Department of Energy[9] | Proton exchange membrane electrolysis | 1600 (in the year 2040) | 1.66 | 74 | 50,000 h | 50−85 | 68 |
| Fuel Cells and Hydrogen Joint Undertaking in Europe[10] | Proton exchange membrane electrolysis | 2500 (in the year 2030) | n/a | n/a | Degradation by 0.12% per 1000 h | n/a | n/a |
| Fuel Cells and Hydrogen Joint Undertaking in Europe[10] | Alkaline water electrolysis | 800 (in the year 2030) | n/a | n/a | Degradation by 0.1% per 1000 h | n/a | n/a |

Note that electrocatalysts play a central role in electrochemical water splitting to reduce electricity consumption and are essential to reach these performance targets. In the past several decades, substantial progress in the development of low-dimensional electrocatalysts has been made, especially in exploring active sites and developing new catalysts.[11-16] These catalysts, however, are commonly studied under laboratory conditions (*e.g.*, with current density of 1−100 mA cm$^{-2}$) and research related to the water splitting mainly focus on fundamental catalyst kinetics.[17] The optimization of a single physical property such as the Gibbs free energy of adsorption for intermediates at a low current density does not usually translate into a good HCD performance because activity and stability of catalysts are also affected by local reaction environment, which is closely related to current density. This fact indicates a large gap between current electrocatalyst studies that focus on low current density conditions and the practical applications where HCD is needed. The research of HCD electrocatalysts is an important aspect in the field of water splitting as it is closely related to the practical applications of this technology. In the past few years, an increasing number of catalysts have been designed and tested, but only few of them deal with industrially-relevant current densities. For example, there are more than 1200 papers in 2014 with the topic of water electrolysis, among them only about 40 papers refer to HCD water electrolysis (Figure 2). Although there are increasing numbers of papers on the development of HCD catalysts, it is still in its infancy as only less than 5% of papers are related to HCDs in all the papers about water electrolysis. Clearly, more efforts should be devoted in HCD considering its important for the large-scale practical implementation of electrolysis.

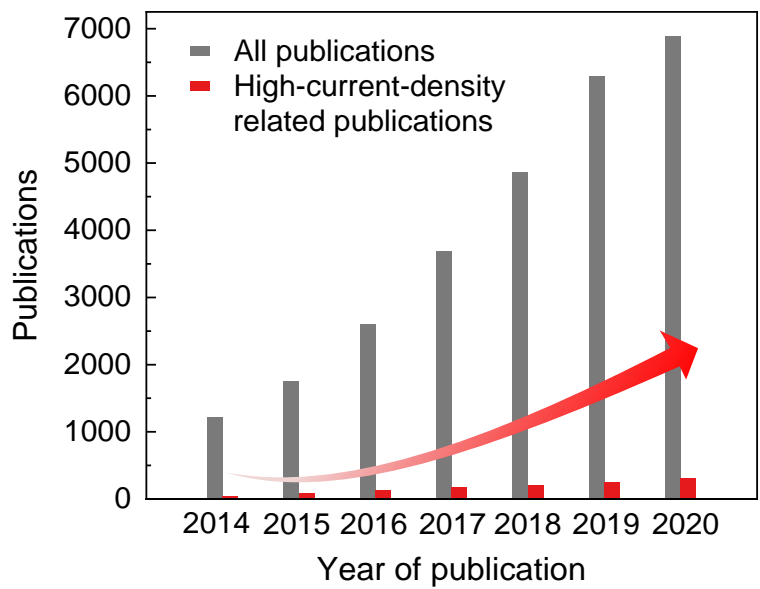

**Figure 2.** Bar chart of the numbers of articles published per year from 2014 to 2020 on electrochemical water splitting (grey bars) and those related to high current density (HCD, red bars) ones. The data for water splitting were obtained by searching the keywords ''("hydrogen evolution") OR ("oxygen evolution") OR ("water splitting")" AND "("electrocataly*") OR ("electrochem*")''. The data for HCDs were obtained by adding ''("high current densit*") OR ("large current densit*") OR ("A cm-2")" as the keywords. All the data were searched in the Web of Science Core Collection.

In the light of recent progress on HCD electrocatalysts, this Review first summarizes the progress in design of HCD electrocatalysts. The catalysts discussed here are those commonly tested at a current density larger than 200 mA cm$^{-2}$ unless point out otherwise. Such a threshold current density is chosen because the operation current density is usually higher than it in industry.[18] Several key aspects that determine the HCD performance of catalysts are discussed, including catalyst dimensionality, surface chemistry, structures, electron transport path, and catalyst-electrolyte interplay, followed by a discussion of the advances and opportunities in the multiscale design of HCD catalysts. Finally, we propose several future directions for research in this important field. This Review mainly focuses on HCD electrocatalysts. Device and system design as well as economic analysis on water electrolysis were well reviewed[19] and are not discussed in detail in this Review.

## 2. Effect of current density on catalytic performance

Catalytic performance is sensitive to local reaction environment, which is current density. Basically, there are two main differences between high and low current density conditions. First, HCD usually means that a large bias is applied to catalysts, leading to an extreme polarization condition far from the equilibrium potential. Second, the electrochemical reaction is violent and fast under HCD conditions, accompanied by the fast consumption of reactants and fast generation of products near the catalyst surface. These differences cause catalytic performance at HCDs different to that at low current densities. The overpotential ($\eta$) of reactions as well as the stability are two main performance metrics, which reflect the effect of HCDs on water splitting. For easier understanding, effect of the HCDs on catalytic performance is compared with low current densities and is discussed from two perspectives of electron transfer and mass transfer (Figure 3). Note that the applied voltage of a half reaction is used instead of current density in some cases since current density is a function of applied voltage. Next, we will show recent progress in understanding how the current density acts on catalytic performance, as a basis of designing electrocatalysts for HCD water splitting.

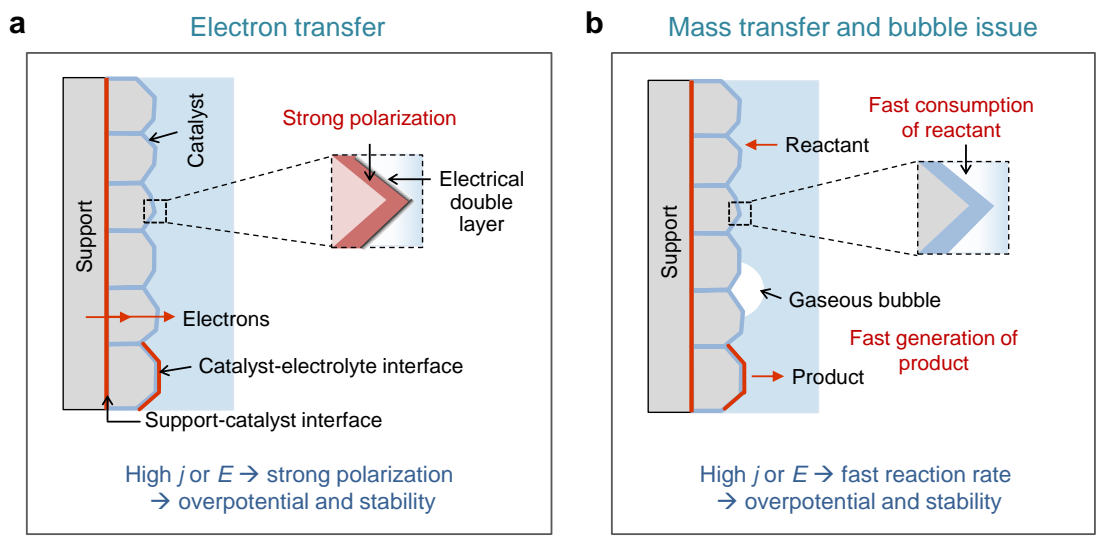

**Figure 3.** Physical models of the electron and mass transfer processes under HCD conditions. Schematics show (a) electron transfer and (b) mass transfer processes at HCDs. Here, support denotes all the materials to load catalytic materials and delivery electrons, and catalyst denotes the catalytically active material or component for reactions.

First, current density or applied voltage affects electron transfer process of reaction that happens at catalyst-electrolyte and catalyst-support interfaces (Figure 3a). At the former interface, catalytic activity is largely determined by the energy needed for adsorption/desorption of intermediates and the rupture/formation of chemical bonds.[20] Recent works show that current densities (or applied voltages of HER or OER reactions) can affect electron-transfer overpotentials by changing the activity of catalysts, such as Pt, IrO$_2$, MoB, and Fe-NiOOH.[21-26] For example, Nong *et al*. find that the current density (or applied voltage) acts on the catalytic activity by charge accumulation in catalysts.[24] In this work, oxidative charges are accumulated and the surface total hole coverage increases with the applied voltage, which are coupled with the electron transfer from IrO$_2$. As a result, the activation free energy for bond formation and rupture decreases linearly with the applied voltage, showing a Tafel slope reducing from 77 mV dec$^{-1}$ to 39 mV dec$^{-1}$ at a voltage up to 1.58 V. This correlation between current density (or applied voltage) and catalytic performance is also shown in HER.[18, 27] For example, Chen *et al.* find that MoB catalyst shows a unduly negative adsorption energy of hydrogen in theory and a slow HER kinetic at low current densities in experiment. [21] However, its catalytic performance surpasses the benchmark Pt catalyst at the current density higher than 250 mA cm$^{-2}$, although MoB and Pt show similar surface areas. Their modelling results attribute such a high activity of MoB at HCDs to the surface hydrogen coverage that increases with current density and tunes the adsorption energy of hydrogen on MoB towards a value even close to zero. The HER activity of Pt, in contrast, decreases as the current density increases. As a result, the MoB needs an overpotential of 334 mV to deliver 1000 mA cm$^{-2}$, while Pt needs around 780 mV to deliver a same current density. He *et al.* show that ultrathin semiconducting catalysts such as $MoS_2$ and $WSe_2$ change their electronic structures and turn into metallic nature by charge accumulation as applied voltage increases, and thus show an increased HER performance.[28] Altogether, these works show that electron transfer process at catalyst-electrolyte interface is affected by current density/applied voltage, which results in a changed catalytic performance at HCDs.

The current density or applied voltage also acts on electron transfer process at the catalyst-support interface. The support is the materials to load catalytic materials and delivery electrons, including the commonly used carbon-based and metal-based materials. Using support materials with a high electrical conductivity and a reduced electrical resistance at catalyst-support interface would decrease the overpotential needed for HCD water splitting. As shown by Zhang *et al.*, $MoS_2$ nanosheets are loaded on different supports and the one on Cu foam shows a better HER performance than those on carbon cloth and Ti foam.[29] $MoS_2$ on Cu foam needs 519 mV to deliver 1000 mA cm$^{-2}$ while that on carbon cloth and Ti foam need 665 mV and 838 mV to deliver the same current density. They find a "soldering effect" between $MoS_2$ and the Cu support, which may reduce the electrical resistance at their interface. Luo *et al.* also use different materials as support and find that NiCoN grown on the Ni foam shows higher HER performance than catalysts on Cu foam, carbon paper, or stainless-steel mat in 1.0 M KOH. To deliver a current density of 100 mA cm$^{-2}$, the best catalyst needs 149 mV.[30] There are some other self-supporting catalysts, similarly, take advantage of small electrical resistance at the catalyst-support interface to achieve an excellent catalytic performance under HCD conditions.[31, 32] All in all, the current density affects electron transfer process happen at the catalyst-electrolyte and catalyst-support interfaces.

Second, current density or applied voltage affects mass transfer process of reaction that happens at the gas-liquid-solid interface containing reactant, product, and catalyst (Figure 3b).[33] The fast consumption of reactants near the catalyst under HCDs may decrease catalytic performance.[34] Liu *et al.* show that concentration of OH- reactant around the tips of $Ni_xFe_{1-x}$ nanocones arrays is high even at HCDs, delivering a current density of 500 mA cm$^{-2}$ at 255 mV for OER in 1.0 M KOH.[35] It should be noted that the formation rate of $H_2$ or $O_2$ bubbles dramatically increases at HCD water splitting and thus hinder reaction process. As shown by some works, thickness of bubble layer increases with the current density and bubbles adhere to catalyst surface cover most of catalyst surfaces and deteriorate their catalytic performance under HCD conditions.[36, 37] Lu *et al.* synthesize $MoS_2$ catalysts with flat and nanostructured morphologies. The latter one shows a superaerophobic nature to the $H_2$ bubbles and thus the sizes of adhesive bubbles are much smaller than those on flat $MoS_2$, resulting in a low overpotential of 500 mV at 170 mA cm$^{-2}$.[38] These works show that current density plays a great impact on mass transfer process and bubble removal on catalysts.

Third, the stability of catalysts, mechanically or chemically, is also influenced by current density.

On the one hand, because bubbles adhesive on catalyst exert a strong interfacial adhesion force on catalyst when they depart from the catalyst, some parts of the catalyst may peel off by the bubbles and deteriorates mechanical stability of catalyst. Such a peeling-off issue of catalyst usually become serious as current density increases unless the interaction force between the catalyst and support is stronger than interfacial adhesion force between the catalyst and bubble. As demonstrated by some binder-free catalysts, they show superior stability under HCD conditions.[39, 40] For example, Zhang *et al*. report a catalyst of CoOOH encapsulated $Ni_2P$ tubular arrays, which shows a stability over 100 h at 1200 mA $cm^{-2}$ for HER.[39] They show that such a catalyst buffers shock of electrolyte convection and hydrogen bubble rupture through release of stress. Despite the progress, the threshold value of interaction force between catalyst and support over which the catalyst will show good robustness under HCDs needs to be found. On the other hand, as HCD means a high electrochemical polarization, the chemical stability may be an issue for HCD water electrolysis. Qin *et al*. study the chemical stability of $Co_xM_{3-x}O$ under HCDs and find that incorporation of Ni in the spinel $Co_3O_4$ improves its long-term chemical stability, while doping of Mn and Ce to spinel $Co_3O_4$ has an opposite effect.[41] As a result, $Co_xNi_{3-x}O_4$ shows the stability over 140 h at 1000 mA $cm^{-2}$ in 1 M KOH for OER. In contrast, other catalysts only run for less than 40 h under the identical conditions. In sum, the effect of current density on mechanical and chemical stability of catalysts should be considered.

## 3. Key aspects need to consider for designing HCD catalysts

Then, we will introduce the recent progress in design of electrocatalysts for HCD water splitting. Five key aspects that are used to engineer catalyst performance at HCDs are summarized and discussed, including catalyst dimensionality, surface chemistry, morphology, electron transport path, and catalyst-electrolyte interplay (Figure 4). Note that some factors that do not affect the performance of catalysts greatly at low current densities become important under HCD conditions. Because some of the factors are also used to engineer the catalyst activity at low current densities, we will focus on the differences of HCDs and avoid discussion about the overlap parts between high and low current densities.

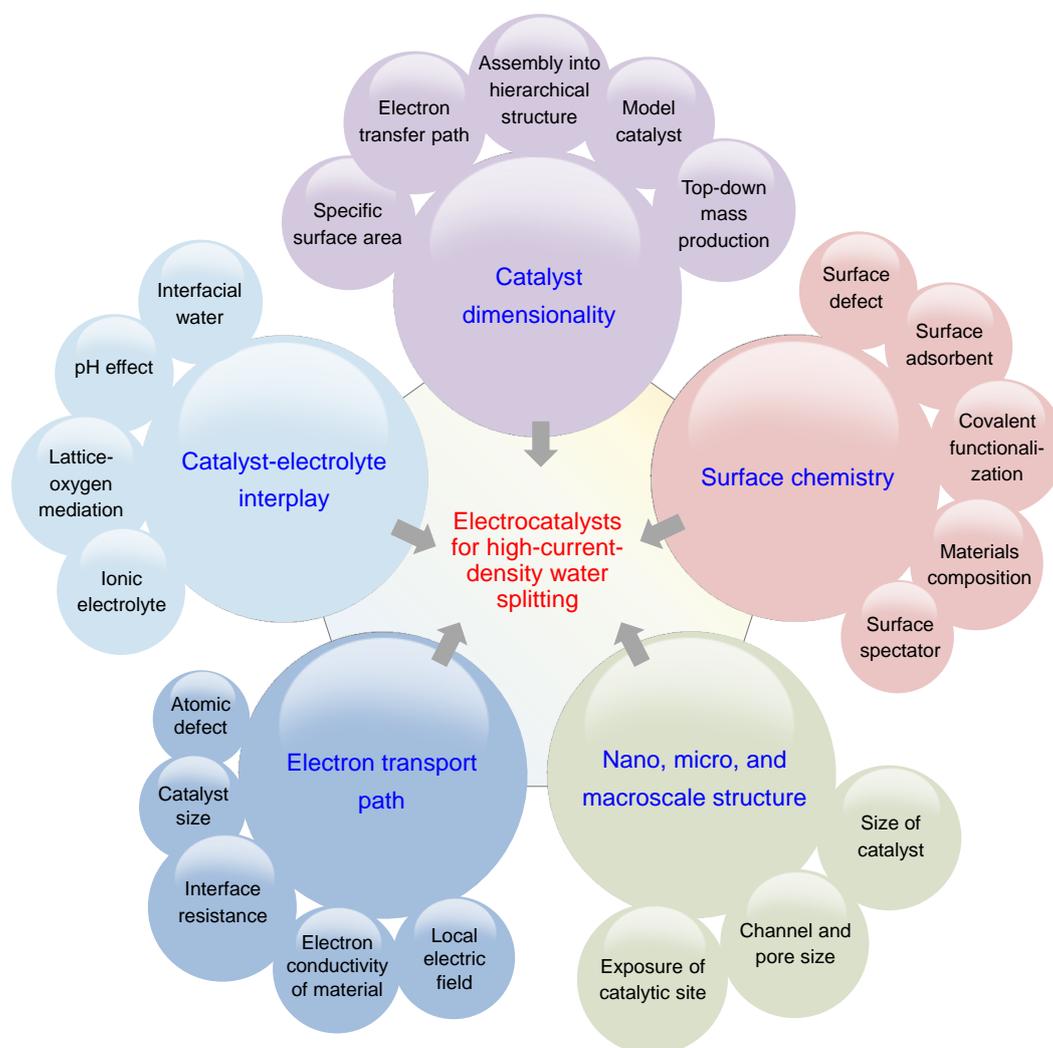

**Figure 4.** Summary of the five key aspects that determine electrocatalyst performance under the HCD conditions, including catalyst dimensionality, surface chemistry, morphology, electron transport path, and catalyst-electrolyte interplay.

## 3.1 Catalyst dimensionality

We shall refer to electrocatalyst particles such as zero-dimensional (0D) single atoms and nanoparticles,

one-dimensional (1D) nanowires and nanotubes, and two-dimensional (2D) nanosheets, as "low-dimensional". The low dimensionality has been reported as an efficient strategy to engineer the HCD performance of catalysts. Electrocatalysts with low dimensionality show many unique points over the bulk catalysts. For example, these materials usually have high specific surface areas, short paths for electron transport in certain directions, variable chemical/physical properties, and the ability to assemble into three-dimensional hierarchical structures. In addition, they show unique properties stem from the dimension effects. 0D nanoparticles can be easily assembled on gas diffusion layers and ion exchange membranes into thin film with large sizes and good uniformity. Besides nanoparticles, other nanomaterials such as single atom, 1D, and 2D catalysts also show their advantages. It is known that single atom catalysts have the highest efficiency of atoms and mass activity for catalysis.[42] Moreover, they can have a high stability by forming strong chemical bonds with the support materials.[43] 1D catalysts such as carbon nanotubes have curved surfaces that can improve the reaction selectivity.[44]. 2D catalysts such as $MoS_2$ can work as model catalysts for mechanism studies because they are atomically flat and have precise structures.[45] Interestingly, they can also be produced from the corresponding bulk materials by top-down methods,[29] making the scaling-up production of 2D catalysts feasible. These properties of low-dimensional catalysts may influence their performance at HCDs by changing the electron transfer and mass transfer processes and thus engineering the dimensionality of catalysts has been used as a strategy of changing catalytic performance at HCDs.

The catalytic performance for HCD water splitting is improved by enhancing electron transfer processes with low dimensionality of catalysts, as shown in Figure 5a. The surface electronic structure of the catalyst and number of coordinated unsaturated atoms in it can be readily engineered by the catalyst particle size, strain, defect, and the ligand effect. Detailly, (i) The low-dimensional catalysts would show fast electron transfer at the catalyst-electrolyte interface and result in a better catalytic activity than bulk materials, and this method uses a consistent idea of increasing catalytic activity at low current densities.[46, 47] (ii) Taking use of low dimensionality to offer a number of active sites that are exposed to reactants, so that the reaction rate increases per electrode area and the HCDs are achieved at a relatively small overpotential. For instance, Bao *et al.* show that $NiCo_2O_4$ ultrathin nanosheets rich in oxygen-vacancy active sites exhibit a current density of 285 mA cm$^{-2}$ at an overpotential of 320 mV, which is superior to the corresponding bulk catalyst and samples with few active sites.[48] (iii) Researchers also find that the low dimensionality of catalysts leads to fast electron transfer at the interface by inducing a spatially inhomogeneous electric field in catalysts, which causes an even strong potential bias at any sharp points on them.[49, 50] (iv) Low-dimensional catalysts may have short paths for electron transport or small resistances at the catalyst-support interface so as to decrease electron transfer overpotential at HCDs.[28, 51] For instance, Liu *et al*. show catalytically active $Co_3O_4$/CoO nanowires that are *in-situ* grown on Cu nanopillars.[52] Such a catalyst shows a relatively lower electron transfer resistance than other control samples, and delivers a current density of 1000 mA cm$^{-2}$ at an overpotential of 391 mV. Altogether, the electron transfer processes at interfaces and inside catalysts are usually greater in low-dimensional catalysts, which guarantee the good catalytic performance for HCD water splitting.

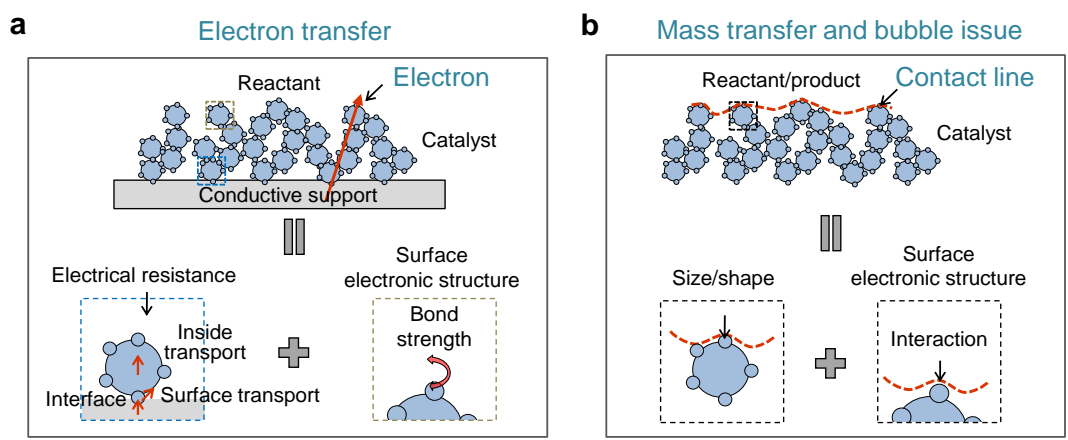

**Figure 5.** Schematics show several main principles that are used to tunes catalytic performance at HCDs by engineering the low dimensionality of catalysts. Modulating catalytic performance at HCDs by (a) electron transfer and (b) mass transfer processes.

The catalytic performance for HCD water splitting is also improved by enhancing mass transfer processes of low-dimensional catalysts, as shown in Figure 5b. (i) The low-dimensional catalysts may induce a changed concentration of reagents/reactants on catalyst,[53] which changes the local chemical potential and improves the catalytic performance at HCDs. For example, Liu *et al*. design a $Ni_xFe_{1-x}$ nanocones array catalyst and compare its catalytic performance at HCDs with those having different tip curvature radii.[35] Their results show that the one with the sharpest tips can concentrate the reactants near the tips and boost its OER performance at HCDs. (ii) Low-dimensional catalyst is efficient to remove $H_2$ and $O_2$ bubbles when they are small because it can reduce the lengths of contact lines at gas-liquid-solid interfaces and thus reduce the interfacial adhesion force between catalysts and bubbles, which benefit catalytic performance at HCDs.[37, 54] For instance, by engineering $MoS_2$ into flat, micro-structured, and nano-structured films, Lu *et al*. show that the nanostructured one shows the highest HER performance at a current density up to 100 mA cm$^{-2}$.[38] On the nanostructured $MoS_2$, $H_2$ bubbles generate and left the catalyst quickly before they grow up to 100 µm in diameter, while bubbles do not leave the flat catalyst unless their diameters are larger than 400 μm. The large numbers of coordinated unsaturated atoms on the surface and the different surface electronic states of low-dimensional catalysts influence their interactions with reactants or products by the electrostatic interaction, van der Waals interaction, hydrogen bonding, *etc*. Such effects directly influence the interfacial adhesion force and thus the mass transfer process. In addition, the mechanical stability of the catalyst related to the removal of bubbles can be improved. The interfacial adhesion force ($F_{int}$) increases with the bubble radius ($r$) and may be larger than the adhesion force of a catalyst film on the support ($F_{ad}$) and therefore rupture the catalyst ($F_{int} > F_{ad}$), unless gas-liquid-solid interfaces ($L(r)$) remains small at HCDs. Specifically, a small bubble with a surface tension ($\sigma$) of ~10 N mm$^{-1}$ generates a small interfacial adhesion force ($F_{int} = \sigma \times L$) of ~$10^0$ N when $L$ is only 0.1 mm, which is smaller than electroplated catalyst films with the $F_{ad}$ of ~$10^0$−$10^2$ N (if $r$ = 0.1 mm).[55] Therefore, the low dimensionality of catalysts has been used to enhance not only the mass transfer efficiency of catalysts but also their mass transfer process-related mechanical stability.

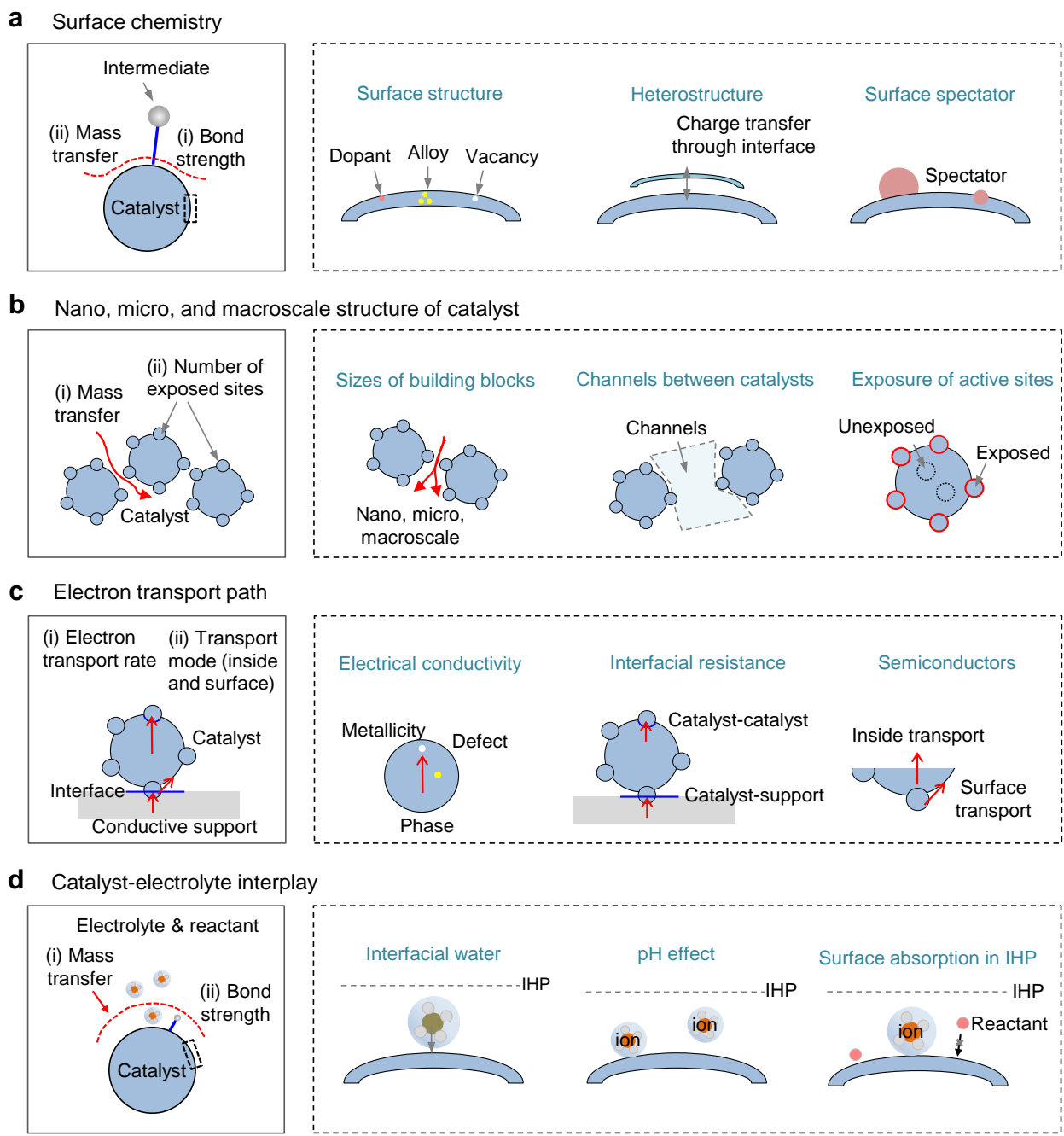

**Figure 6.** Schematics showing several key aspects of design of catalysts for HCD water splitting, including (a) catalyst surface chemistry, (b) catalyst morphology, (c) electron transport path, and (d) catalyst-electrolyte interplay, respectively.

## 3.2 Surface chemistry

Engineering surface chemistry of catalysts can tune catalyst performance for HCD water splitting by

(1) changing the bond strength between intermediates and catalytic sites and (2) improving the mass transfer efficiency of catalysts (Figure 6a). Indeed, catalyst surface (*e.g.*, composition, structure, defect, strain, doping), heterostructure with interfacial interaction, and surface spectator, *etc.*, all influence the surface chemistry of the catalysts and some of them have been used to tune HCD performance.[56-60] Engineering the surface chemistry of catalysts is a basis to achieve high performance at HCDs.

Engineering catalyst surface changes the surface electronic structure and the bond strength between intermediates and active sites. For example, Luo *et al*. design a catalyst that composes of the $MoS_2$ nanosheets with $Mo_2C$ nanoparticles decorated on their edges ($MoS_2/Mo_2C$), which shows a superior performance at HCDs as compared with the pure $MoS_2$ catalyst (Figure 7a).[18] Specifically, it shows overpotentials of 227 mV in acidic medium and 220 mV in alkaline medium at the current density of 1000 mA $cm^{-2}$. Although $MoS_2$ and $MoS_2/Mo_2C$ catalysts show the same microscopic morphology, they have different surface chemistry compositions (Figure 7b). The latter one shows fast HER kinetic due to surface oxygen groups formed on $Mo_2C$ during the HER process, which decrease the energy barriers for both adsorption/desorption of hydrogen and dissociation of water at HCDs. Moreover, they find that such a difference in surface chemistry makes the bubble release easily on $MoS_2/Mo_2C$ catalyst and fast mass transfer efficiency. In another example, Zheng *et al*. designed a $MoS_2$ nanofoam catalyst co-confining selenium in the surface and cobalt in inner layers (Figures 7c-d).[61] By engineering the surface chemistry of $MoS_2$, the catalyst needs an overpotential of 382 mV to deliver a current density of 1000 mA $cm^{-2}$ (Figure 7e) and its performance is stable for 360 h without decay. The theoretical results attribute the HCD performance to the optimized hydrogen adsorption activity of both in-plane and edge active sites on the doped $MoS_2$. Recently, single atom catalysts loaded on proper supports are also used for HCD water splitting due to their advantages of high atom utilization efficiency and high activity.[62, 63] For example, Liu *et al*. show that by loading 0.49 wt% Pt single atoms on RuCeO$_x$, the catalyst shows an overpotential of 320 mV at 600 mA cm$^{-2}$ and is better than the commercial 20 wt% Pt/C catalyst.[64]

Constructing heterostructures with charge transfer through the interfaces of each components is another strategy to tune catalyst performance at HCDs. For example, Yao *et al*. report that a graphdiyne/molybdenum oxide (GDY/MoO$_3$) catalyst shows a *sp* C−O−Mo hybridization on the interface between graphdiyne and molybdenum oxide (Figures 7f-g), which facilitates the charge transfer and boosts the dissociation process of H$_2$O molecule.[32] The heterostructure catalyst exhibits a current density of 1200 mA cm$^{-2}$ at an overpotential of ~1850 mV in 0.1 M KOH, which exceeds both the HER performance of graphdiyne and molybdenum oxide catalysts without such an interfacial interaction (Figure 7h).

In addition to the surface composition of the catalyst and heterostructure structure, surface spectators that are chemically bonded to or form a composite/heterostructure with the catalyst also affect HCD performance. Their functions include, but are not limited to, the destabilization of reactant species, and changing the bond strength between intermediates and catalytic sites,[65] despite that the spectators may be not catalytic sites in the catalytic materials. All the progress show that surface chemistry engineering is an effective strategy for designing catalysts for HCD water splitting.

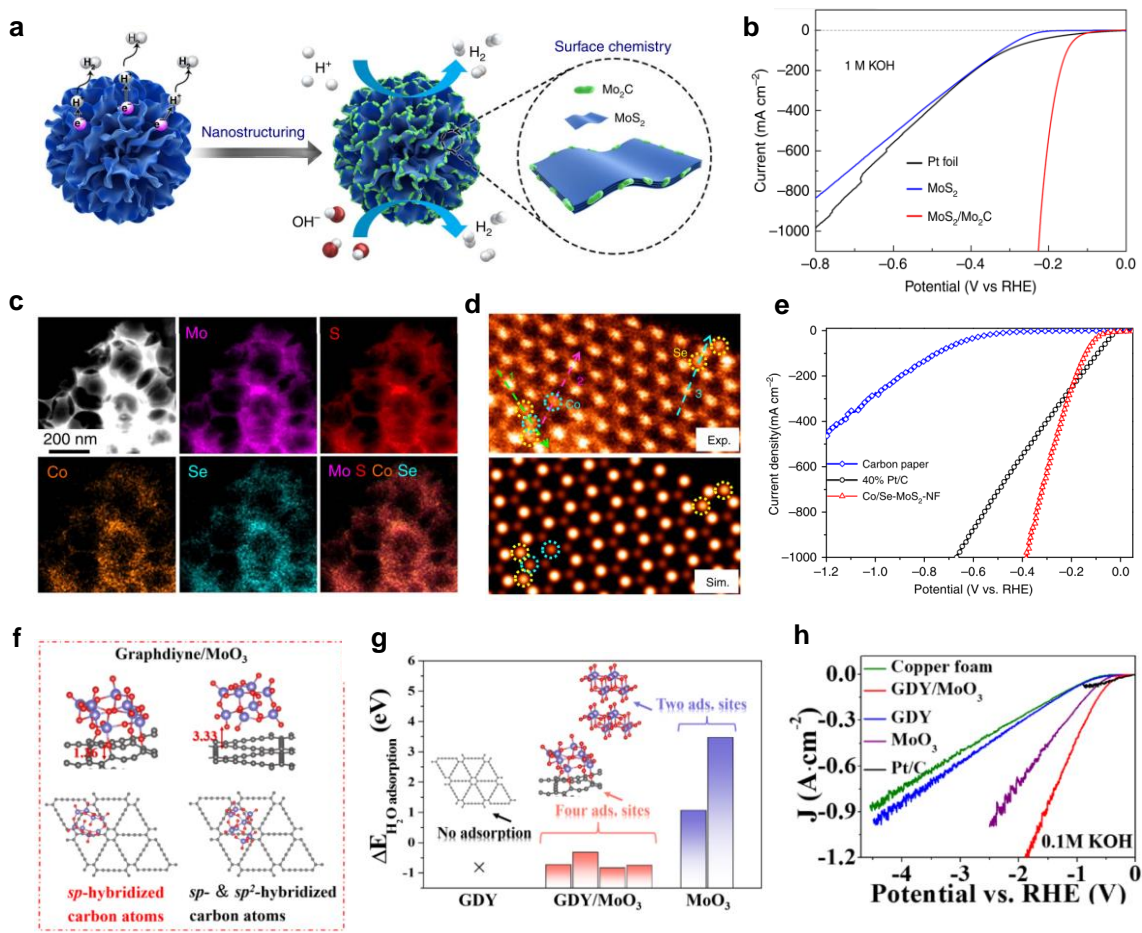

**Figure 7.** Engineering the surface chemistry of catalysts for HCD water splitting. (a-b) MoS₂/Mo₂C catalyst modified by surface oxygen. Reproduced from ref. [18] with permission from the Springer Nature, copyright 2019. (c-e) MoS₂ catalyst co-confining selenium in the surface and cobalt in inner layers. Reproduced from ref. [61] with permission from the Springer Nature, copyright 2020. (f-h) graphdiyne/molybdenum oxide heterostructure showing a *sp* C−O−Mo hybridization on the interface between graphdiyne and molybdenum oxide. Reproduced from ref. [32] with permission from the American Chemical Society, copyright 2021.

### 3.3 Nano, micro, and macroscale structure of catalyst

Because the catalyst is used in the form of a film that has to contact a current collector (or support), each unit of low-dimensional catalysts (the building blocks of the film) need to assemble in a certain way. The nano, micro, and macroscale structure of such a catalyst film jointly determine the size of the channels and the exposure of active sites to the electrolyte (Figure 6b). This is also an aspect that can be engineered towards better HCD performance for water splitting by (1) increasing mass transfer ability and (2) increasing the numbers of active sites for catalysis. Note that the effect of the sizes of the building blocks on HCD performance have been discussed in the part 3.1 "catalyst dimensionality".

First, the channels (pores) constructed by catalytic building blocks are engineered to ensure smooth electrolyte supply and bubble removal, and thus decrease the mass transfer overpotential for HCD water splitting.[66-68] For example, Park et al. show that the CuCo-oxide OER catalyst grown on a nickel foam delivers a current density of 2200 mA cm$^{-2}$ at a cell voltage of 1.9 V using Pt/C as HER catalyst (Figure 8a).[69] They use their catalyst in an anion exchange membrane water electrolyzer and compare the mass transfer overpotential with the IrO$_2$ catalyst. At a current density larger than 200 mA cm$^{-2}$, the mass transfer overpotential contributes greatly to the total overpotential, which accounts for 10.4% on CuCo-oxide catalyst while 35.4% on IrO$_2$ catalyst at 1200 mA cm$^{-2}$ (Figure 8b). The CuCo-oxide catalyst shows a much smaller mass transfer overpotential due to its highly porous structure constructed by catalytic building blocks that ensure smooth supply of electrolyte at HCDs (Figure 8c). Yu et al. report a catalyst film consisting of Ni$_2$P nanowire arrays grown on a Ni foam (Figures 8d-f), which enables the fast release of H$_2$ bubbles in the HER at HCDs and delivers 1000 mA cm$^{-2}$ at an overpotential of 306 mV (Figure 8g).[70] Despite decade of efforts, more quantitative understanding of the effects of channel or pore sizes on mass transfer and catalyst performance is still needed.

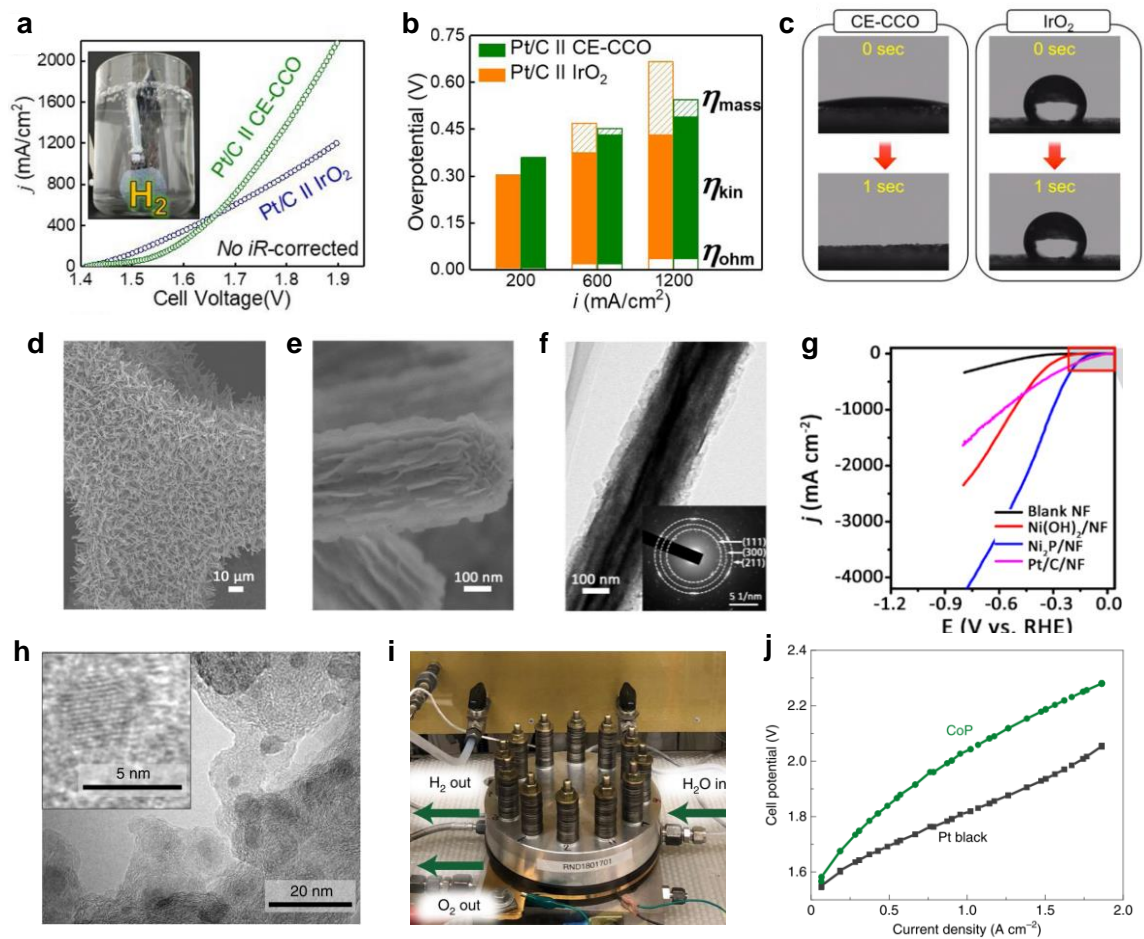

**Figure 8.** Engineering the nano, micro, and macroscale structure of catalysts for HCD water splitting. (a-c) CuCo-oxide catalyst showing smaller mass transfer overpotential than the IrO$_2$ at HCDs. Reproduced from ref. [69] with permission from the Elsevier, copyright 2020. (d-g) Ni$_2$P nanowire arrays grown on a Ni foam promoting H$_2$ bubble removal for HER at HCDs. Reproduced from ref. [70] with permission from the American Chemical Society, copyright 2019. (h-j) CoP nanoparticles on the surface of a high-surface-area carbon support for HER at HCDs. Reproduced from ref. [71] with permission from the Spring Nature, copyright 2019.

Second, the exposure of catalytic sites to the electrolyte ensures efficient uses of the catalysts and enables the availability of many active sites, with which catalyst performance at HCDs increases. For example, King *et al*. report a catalyst composed of CoP nanoparticles on the surface of a high-surface-area carbon support that is integrated into an 86 cm$^2$ PEM electrolyser (Figures 9h-i).[71] Despite its lower mass activity than Pt particles, it shows a good apparent current density due to the large loading of CoP, which operates at 1860 mA cm$^{-2}$ for >1700 h (at 50 °C and 400 psi, pounds per square inch) of continuous hydrogen production (Figure 8j). Engineering the exposure of catalytic sites on conductive supports is a common strategy that have been widely used. The role of nano, micro, macroscale structures of the catalyst in the HCD electrocatalysis involves controllable assembly of low-dimensional building blocks into hierarchical catalysts needs more study. To date, various catalyst/support structures have been synthesized but there are still few methods for scaling-up the production of electrocatalysts with many active sites. The progress in engineering nano, micro, and macroscale structure of catalysts shows the effectiveness of this strategy for designing HCD catalysts.

## 3.4 Path for electron transport

Constructing electron transport path to reduce electrical resistance is another strategy to obtain good HCD performance of catalysts by (1) changing the rate of electron transport in the catalyst and (2) changing the electron transfer mode from the conductive support to the catalyst-electrolyte interface (Figure 6c). To reduce these electrical resistances, researchers focus on several strategies that change electron transport from support to catalytic sites, including electron transport in a conductive catalyst, electron transfer at catalyst-support and catalyst-catalyst interfaces, and electron transport mode in a semiconducting catalyst.

Using materials with high electrical conductivity as the catalysts or supports is a general idea to improve catalytic performance at HCDs. The materials include carbon, metals, and some metallic compounds. There are several ways to change the conductivity of materials, such as phase engineering and defect engineering. A main challenge is to take use of high conductivity and meanwhile to get high intrinsic activity. Recently, Yang *et al*. directly grow a metallic 2H-phase $Nb_{1.35}S_2$ catalyst on a highly-conductive glassy carbon support (Figure 9a), where it shows a conductivity of $10^3$ S $cm^{-1}$.[72] This conductivity is comparable to the bulk $3R-NbS_2$ and an order of magnitude lower than Pt metal. Interestingly, the intrinsic activity of $Nb_{1.35}S_2$ catalyst is higher than bulk $3R-NbS_2$ and is comparable to that of Pt wire. As a result, it has an excellent HCD performance with an overpotential of ~370 mV at a current density of 5000 mA $cm^{-2}$ normalized by the projected surface area of catalyst for the HER (Figures 9b-c). For OER catalysts, due to the oxidation potentials usually convert metallic materials into (hydro)oxides, the conductivity of catalysts and their performance at HCDs may be changed.[73] It is reported that a proper content of defects or vacancies in metal oxides increase their electrical conductivities. For example, Bao *et al*. show that rich oxygen vacancies in ultrathin $NiCo_2O_4$ nanosheets promote its OER reactivity and the delocalized electrons around the oxygen vacancy are easy to be excited to the conduction band to enhance the conductivity of catalyst.[48] The catalyst therefore delivers a current density of 285 mA $cm^{-2}$ at an overpotential of 320 mV. Besides, despite many studies on the design of electrocatalysts using materials with good conductivity, it is still not known that whether the better electrical conductivity the better HCD performance or there is a critical value of the electrical conductivity above which a higher conductivity does not result in a better HCD performance.

Boosting electron transfer through catalyst-support and catalyst-catalyst interfaces directly improves the catalyst performance for HCD water splitting. Indeed, the interfacial resistance is smaller between a metallic catalyst and a conductive support than that of semiconducting or insulating catalyst. Based on this, self-supporting catalysts/electrodes that use a support such as metal foam and carbon material with catalytic materials directly grown on its surface are intensively studied.[74, 75] Coupling effects between a catalyst and its support may inject or withdraw electrons from the catalyst, and thus change the interfacial resistance.[45, 76] To reduce this resistance, the thickness or size of the catalysts is crucial, especially for semiconducting catalysts.[28, 77] Moreover, the electron transport paths should be short, which requires a small thickness/size of the catalyst building blocks and their proper stacking. For example, Yan *et al.* show that interface between electrodeposited catalyst materials (cerium dioxide and nickel hydroxide) and support materials (graphite with nitrate inserted into its layers) becomes strong due to charge transfer between them, which results in a highly porous and high-loading film via proper stacking.[75] The catalyst exhibits an overpotential of 310 mV at 1000 mAcm$^{-2}$ and a durability over 300 h.

Some semiconducting materials are also used for HCD water splitting and their electron transport mode is found to be interesting.[13] Usually, a semiconducting catalyst needs to be combined with a conductive support during use and thus show a three-phase electrochemical interface that consists of a catalyst, a conductive support, and the electrolyte. The electron transfer mode is different to that for metallic catalysts. Recent studies have shown that electrons transfer directly from the conductive supports to the catalyst surface along the three-phase contact lines.[51, 78] He *et al.* find that the surface of 2D MoS$_2$ nanosheets even shows a metallic state as current density/applied voltage increase while the bulk region remains semiconducting (Figures 9d-e).[28] Altogether, these results indicate that constructing efficient electron transfer path is necessary to achieve good catalytic performance for HCD water splitting.

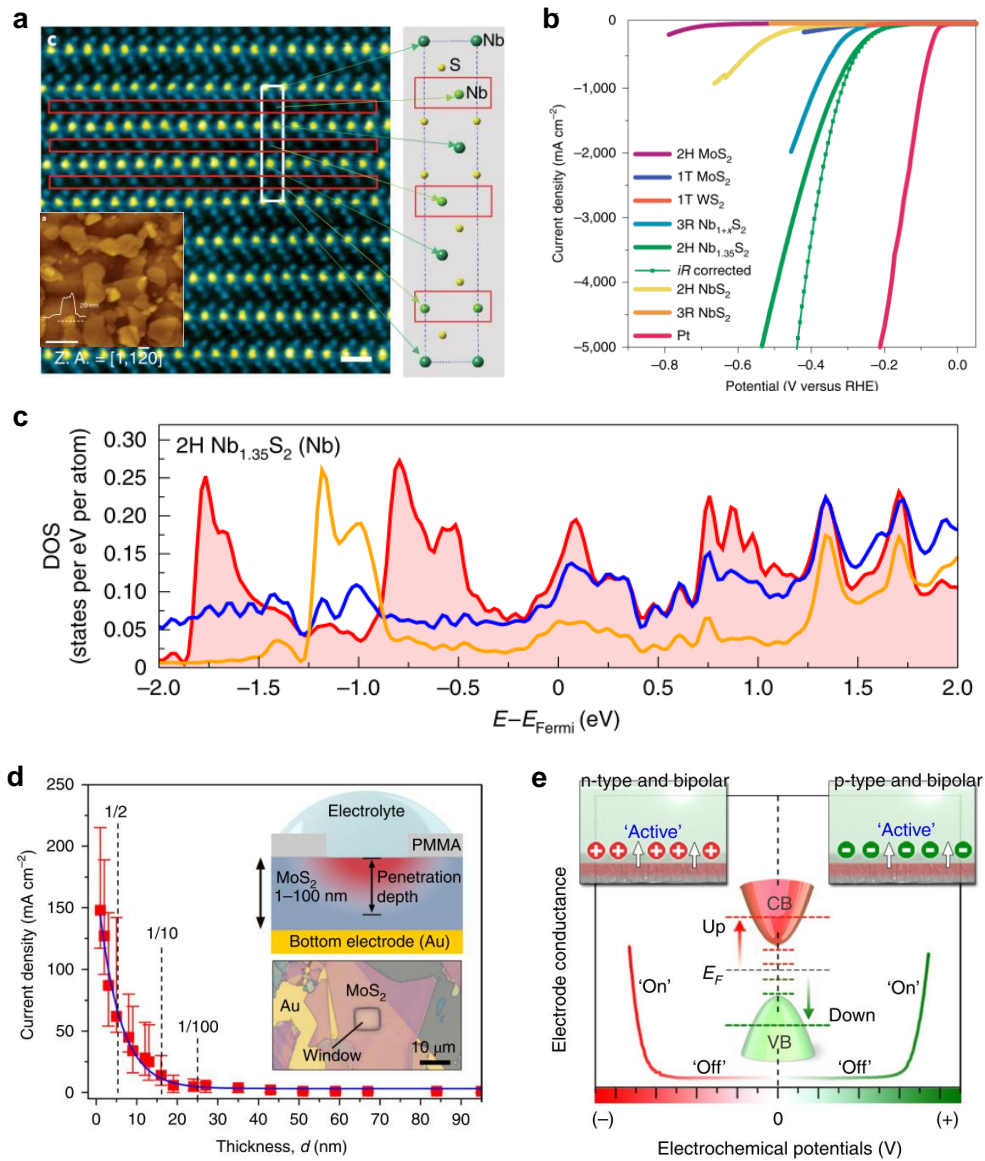

**Figure 9.** Engineering the electron transport paths of catalysts for HCD water splitting. (a-c) Metallic 2H-phase Nb$_{1.35}$S$_2$ catalyst with a high electrical conductivity grown on a glassy carbon support. Reproduced from ref. [72] with permission from the Spring Nature, copyright 2019. (d-e) 2D MoS$_2$ nanosheets showing a semiconducting-to-metallic state transition at HCDs. Reproduced from ref. [28] with permission from the Spring Nature, copyright 2020.

### 3.5 Catalyst-electrolyte interplay

The interaction between the catalyst and the electrolyte (or reactant) has received great attention in recent years because it affects our fundamental understanding of the catalytic sites in operating conditions. The HCD performance of catalysts is tuned by engineering catalyst-electrolyte interplay by (1) changing bond strength between intermediates and active sites, and (2) changing mass transfer efficiency (Figure 6d). Different to surface adsorbents, the species involved in the catalyst-electrolyte interplay directly derived from the electrolytes or reactants during reactions. Several strategies have been developed to take use of catalyst-electrolyte interplay towards better HCD performance, including engineering the interaction between catalyst and interfacial water in inner Helmholtz plane (IHP), engineering the adsorbents in the IHP, taking use of the effect of pH values, and so on.[23, 79, 80]

First, the interplay between catalyst and interfacial water in the IHP changes not only the bond strength between intermediates and the catalyst but also the concentration of protons, and hence tunes the HCD performance of the catalyst. For example, Ledezma-Yanez $et$ $al$. report that adding nickel to a Pt(111) surface accelerates the reaction rate of HER in alkaline media.[81] They attribute the different activities of Pt(111) and nickel decorated Pt(111) to the reorganization of the interfacial water that accommodates charge transfer through the electric double layer. The energetics are controlled by the strength of the interaction between water molecules and the interfacial field. In another work, Jin $et$ $al$. reported that Ni-SN@C would facilitate water adsorption and weaken hydrogen adsorption, leading to the generation of hydronium ions near the surface of the catalyst in a high-pH electrolyte. In contrast, on the Ni@C and $Ni_3N$ catalysts, the H* would be directly converted to hydrogen molecule instead of forming hydronium as the intermediate at HCDs. The interaction between catalyst and interfacial water in the IHP can influence the reaction mechanism and activities.

Second, adsorbents in the IHP on catalyst surface affect the rupture/formation of chemical bonds, stabilization of intermediates, $etc$. Their roles in the catalytic performance have recently aroused interest.[18, 46, 82] For example, Zhang *et al*. construct a highly conductive edge-enriched $Ni_{0.2}Mo_{0.8}N/Ni$ hybrid catalyst, which delivers 300 mA cm$^{-2}$ at an overpotential of 70 mV for HER in 1 M KOH.[80] This good HCD performance is attributed to tip-enhanced-like local electric field around the topmost Ni nanoparticles, leading to an increased concentration of K ions in the IHP. The types of surface adsorbents may be affected by the applied current densities or overpotentials, resulting in a different catalytic performance at HCDs. Very recently, Nong *et al*. find that the coverage of holes is changed with a change in overpotential and is coupled with deprotonation of the $IrO_2$ catalyst for OER.[24] The local adsorption of a high concentration of reagents on a catalyst during reactions changes the chemical potential and hence HCD performance.[53] Engineering the coverages or types of adsorbents on catalysts for HCD water splitting deserves more attention in future.

Third, the pH value of the electrolyte is found to influence the HCD performance of catalysts.[23] For example, Luo *et al*. show that $Mo_2C$ nanoparticle catalyst is modified by different types of surface oxygen species as pH value changes.[18] Specifically, $O_x$-group is prone to modify $Mo_2C$ surface when pH value is high while OH-group is found at surface at low pH conditions. The former one shows a relatively low energy barrier for water dissociation than unmodified $Mo_2C$, and the latter one shows a low energy for adsorption/desorption of hydrogen. As a result, such a catalyst exhibits a current density of 1000 mA cm$^{-2}$ at overpotentials of 220 mV and 227 mV in electrolytes of pH = 14 and pH = 1, respectively. Note that the studies of the roles of catalyst-electrolyte interplay in HCD electrocatalysis need to be combined with advanced *in-situ* spectroscopy characterization to help researchers to understand HCD catalysis in real operating conditions.

## 4. Multiscale design of catalysts toward HCD electrocatalysis

The five aspects discussed above are used to engineer catalysts to achieve efficient HCD water splitting.

The engineering of any single aspect cannot produce catalysts with a superior HCD performance, and hence a multiscale design strategy for catalysts that engineering several aspects at the same time is necessary. This design strategy has recently been used to achieve high-performance catalysts under the HCD conditions.[40, 83-88] For example, Yang *et al.* recently show that a Pt/Ni-Mo electrocatalyst only needs an overpotential of 113 mV to reach an ultrahigh current density of 2000 mA cm$^{-2}$ in the saline-alkaline electrolyte and could run at 2000 mA cm$^{-2}$ for 140 h without performance decay.[89] Such an excellent performance at HCDs can be attributed the multiscale design strategy of catalyst that taking consideration of surface chemistry of catalyst, electron transport pathway, and catalyst morphology in the design. Despite progress has been made in the multiscale design strategy of catalysts for HCD water splitting, there are some points also need to be developed.

To optimize the multiscale design of catalysts for HCD electrocatalysis, the relations between the five factors need to be comprehensively considered because there may be many trade-offs. Here are some examples. A low-dimensional catalyst has a large number of exposed catalytic sites, which however may lower the rate of mass transfer because the effective diffusion length may be increased due to their higher sinuosity. The catalyst film assembled by low-dimensional catalyst may also decrease the electrical conductance of the film by introducing too many catalyst-catalyst interfaces. The small sizes of low-dimensional catalysts may also reduce their stability due to a numbers of surface unsaturated atoms. Presence of the surface adsorbents and spectators on catalysts may destabilize the reactants and stabilize the intermediate to produce a higher intrinsic activity, however, they may also cover some catalytic sites giving a lower electrochemical surface area and producing interfacial resistance. A catalyst with an ultrahigh surface area has a large number of catalytic sites, but this may increase the transport resistance of gas or ions at HCDs. Based on the above discussions, the multiscale design strategy of catalysts for HCD water splitting is complicated in practice because some relations between the main factors need to be balanced. There is still a plenty room for the further development of the multiscale design of HCD electrocatalysts.

## 5. Conclusions and perspectives

In this review, the effect of HCD conditions on local reaction environment and catalytic performance has been discussed firstly. By discussing recent advances in HCD electrocatalysts, several key aspects that have been used for engineering catalysts toward efficient HCD water splitting have been then summarized, including low dimensionality of catalysts, surface chemistry electron transport path, morphology, and catalyst-electrolyte interplay. Finally, we highlight the multiscale design strategy of efficient HCD catalysts. The performance of state-of-the-art HCD electrocatalysts is still far from their target values. Therefore, it is still crucial to explore HCD electrocatalysts for the industrial applications. Several research directions should be pursued in future.

**(1) In-depth understanding about electrochemical interfaces under HCD conditions.**

The local electrochemical environment at electrochemical interface is greatly affected by current densities. However, mechanism understanding about electrochemical interface under HCD conditions is still limited, hindering the rational design of high-performance electrocatalysts for water electrolysis working at industrially-relevant HCDs. *In-situ* spectroscopic characterization and other advanced characterization techniques needed to be developed to understand how the HCD affect the electrochemical interfaces and processes, which however, are very challenging as generation of gas bubbles is violent at HCDs and they may scatter rays. Besides, theoretical methods that are suitable for HCD conditions need to be developed.

**(2) Developing multiscale design strategies for HCD electrocatalysts.**

The performance of HCD electrocatalysts is determined by several aspects at atomic, nano-meter, and micro-meter scales. To obtain efficient and stable HCD catalysts, multiscale design strategies need to be developed that consider all the aspects at the same time. And the trade-off in activity and stability of HCD catalysts caused by these aspects need systematic studies. It is crucial to explore strategies for OER catalysts at HCDs that show smaller overpotentials than target values, such as the middle-term target values for alkaline OER is 1.43 V at 500 mA cm$^{-2}$. Currently, excellent performance at HCDs have been shown by some "self-supporting" catalysts in alkaline electrolyzers and H-cells, where the catalytic materials are grown on porous and conductive substrates.[85, 89] Great efforts are expected to develop multiscale design strategies for HCD catalysts suitable for electrolyzers based on ion exchange membranes. Stability of OER catalysts under acid environment at HCDs is also important.

**(3) Developing standards for evaluating catalyst performance relevant to industrial use.**

This requirement calls for the benchmarking and assessment of catalytic performance using standardized materials and conditions, and moreover, under industry-relevant conditions.[19, 90] The key performance metrics need to be identified to assess the catalyst performance under HCDs. Moreover, electrolyzers with similar configurations used in practical applications should be used for assessment of HCD catalysts. In addition, more efficient and time-saving stability and durability test methods are needed.

**(4) Economic considerations for the future development of HCD electrocatalysts.**

Beyond these scientific and technological issues, the economic considerations are necessary for the future development of HCD catalysts. To achieve large-scale implementation of HCD reactions, green productions and costs of the technologies, including raw materials/chemicals, production costs of catalysts, and the energy consumption during electrochemical reactions, need to be considered.

We project continued endeavors towards the development of efficient catalysts for the electrochemical water splitting technologies and their wide implementation under the HCD conditions would make them gaming-changing players toward global carbon neutrality and sustainable development future.

**Acknowledgement.**

We thank Prof. Hui-Ming Cheng, Fengning Yang, and Heming Liu for helpful discussions. We also acknowledge support from the National Natural Science Foundation of China (No. 51722206), Guangdong Innovative and Entrepreneurial Research Team Program (No. 2017ZT07C341), the Bureau of Industry and Information Technology of Shenzhen for the "2017 Graphene Manufacturing Innovation Center Project" (No. 201901171523), and the Shenzhen Basic Research Project (No. JCYJ20200109144620815).

& Sons, Inc., New York 2001.